\documentclass[prl,a4paper,bm,twocolumn,showpacs,superscriptaddress]{revtex4}
\usepackage{bm}
\usepackage{amssymb}
\usepackage{amsfonts}
\usepackage{amsmath}
\usepackage{graphicx}
\usepackage[dvips]{color} 

\definecolor{g-blue}{rgb}{0.83,0.95,1}
\definecolor{Blue}{rgb}{0.5,0.5,1}
\definecolor{DarkBlue}{rgb}{0.00,0.00,0.58}
\definecolor{g-yellow}{rgb}{1,1,0.7}
\definecolor{g-green}{rgb}{0.9,1,0.9}
\definecolor{green}{rgb}{0,0.6,0}
\definecolor{Green}{rgb}{0,0.4,0}
\definecolor{cyan}{rgb}{0,0.7,0.7}
\definecolor{black}{rgb}{0,0,0}
\definecolor{grey}{rgb}{0.4 ,0.4 ,0.4 }

\begin{document}

\def\Fbox#1{\vskip1ex\hbox to 8.5cm{\hfil\fboxsep0.3cm\fbox{%
  \parbox{8.0cm}{#1}}\hfil}\vskip1ex\noindent}  


\newcommand{\eq}[1]{(\ref{#1})}
\newcommand{\Eq}[1]{Eq.~(\ref{#1})}
\newcommand{\Eqs}[1]{Eqs.~(\ref{#1})}
\newcommand{\Fig}[1]{Fig.~\ref{#1}}
\newcommand{\Figs}[1]{Figs.~\ref{#1}}
\newcommand{\Sec}[1]{Sec.~\ref{#1}}
\newcommand{\Secs}[1]{Secs.~\ref{#1}}
\newcommand{\Ref}[1]{Ref.~\cite{#1}}
\newcommand{\Refs}[1]{Refs.~\cite{#1}}

\def\be{\begin{equation}}\def\ee{\end{equation}}
\def\bea{\begin{eqnarray}}\def\eea{\end{eqnarray}}
\def\bse{\begin{subequations}}\def\ese{\end{subequations}}
\newcommand{\BE}[1]{\begin{equation}\label{#1}}
\newcommand{\BEA}[1]{\begin{eqnarray}\label{#1}}
\newcommand{\BSE}[1]{\begin{subequations}\label{#1}}

\let \nn  \nonumber  \newcommand{\br}{\\ \nn}
\newcommand{\BR}[1]{\\ \label{#1}}
\def\hf{\frac{1}{2}}
\let \= \equiv \let\*\cdot \let\~\widetilde \let\^\widehat \let\-\overline
\let\p\partial \def\pp {\perp} \def\pl {\parallel}
\def\ort#1{\^{\bf{#1}}}
\def\Trans{^{\scriptscriptstyle{\rm T}}}
\def\x{\ort x} \def\y{\ort y} \def\z{\ort z}
 \def\bn{\bm\nabla} \def\1{\bm1} \def\Tr{{\rm Tr}}
\def\Re{\mbox{  Re}}
\def\<{\left\langle}    \def\>{\right\rangle}
\def\({\left(}          \def\){\right)}
 \def \[ {\left [} \def \] {\right ]}

\renewcommand{\a}{\alpha}\renewcommand{\b}{\beta}\newcommand{\g}{\gamma}
\newcommand{\G} {\Gamma}\renewcommand{\d}{\delta}
\newcommand{\D}{\Delta}\newcommand{\e}{\epsilon}\newcommand{\ve}{\varepsilon}
\newcommand{\E}{\Epsilon}\renewcommand{\o}{\omega} \renewcommand{\O}{\Omega}
\renewcommand{\L}{\Lambda}\renewcommand{\l}{\lambda}
\renewcommand{\t}{\tau}
\def\r{\rho}\def\k{\kappa}
\def\t{\theta } \def\T{\Theta } \def\s{\sigma} \def\S{\Sigma}

\newcommand{\B}[1]{{\bm{#1}}}
\newcommand{\C}[1]{{\mathcal{#1}}}    
\newcommand{\BC}[1]{\bm{\mathcal{#1}}}
\newcommand{\F}[1]{{\mathfrak{#1}}}
\newcommand{\BF}[1]{{\bm{\F {#1}}}}

\renewcommand{\sb}[1]{_{\text {#1}}}  
\renewcommand{\sp}[1]{^{\text {#1}}}  
\newcommand{\Sp}[1]{^{^{\text {#1}}}} 
\def\Sb#1{_{\scriptscriptstyle\rm{#1}}}

\title{Superfluid vortex front at $T\rightarrow 0$: Decoupling  from the reference frame}

\author{J.J.~Hosio}
\affiliation{Low Temperature Laboratory, School of Science, Aalto University, FI-00076 AALTO, Finland}

\author{V.B.~Eltsov}
\affiliation{Low Temperature Laboratory, School of Science, Aalto University, FI-00076 AALTO,   Finland}

\author{R.~de~Graaf}
\affiliation{Low Temperature Laboratory, School of Science, Aalto University, FI-00076 AALTO, Finland}

\author{P.J. Heikkinen}
\affiliation{Low Temperature Laboratory, School of Science, Aalto University, FI-00076 AALTO, Finland}

\author{R.~H\"anninen}
\affiliation{Low Temperature Laboratory, School of Science, Aalto University, FI-00076 AALTO, Finland}

\author{M.~Krusius}
\affiliation{Low Temperature Laboratory, School of Science, Aalto University, FI-00076 AALTO,
Finland}

\author{V.S.~L'vov}
\affiliation{Department of Chemical Physics, The Weizmann Institute of
  Science, Rehovot 76100, Israel}

\author{G.E.~Volovik}
\affiliation{Low Temperature Laboratory, School of Science, Aalto University, FI-00076 AALTO,   Finland}
\affiliation{Landau Institute for Theoretical Physics,
Kosygina 2, 119334 Moscow,  Russia}

\date{\today}

\begin{abstract}
  Steady-state turbulent motion is created in superfluid $^3$He-B at low
  temperatures in the form of a turbulent vortex front, which moves
  axially along a rotating cylindrical container of $^3$He-B and replaces
  vortex-free flow with vortex lines at constant density. We present the
  first measurements on the thermal signal from dissipation as a function
  of time, recorded at $0.2 \, T_\mathrm{c}$ during the front motion, which
  is monitored using NMR techniques. Both the measurements and the 
  numerical calculations of the vortex dynamics show that at low
  temperatures the density of the propagating vortices falls well below
  the equilibrium value, i.e. the superfluid rotates at a smaller angular
  velocity than the container. This is the first evidence for the
  decoupling of the superfluid from the container reference frame in the
  zero-temperature limit.
\end{abstract}
%

\maketitle 

\textbf{Introduction:}--For eighty years the phenomenological two-fluid
model of Tisza and Landau has been the starting point for discussions on
dynamics in superfluids. Onsager and Feynman introduced the
quantized vortex, which plays an all important role in the dynamics:
Vortices mediate the interaction between the
superfluid and normal components of the liquid via
reactive and dissipative friction forces \cite{Volovik2003}.
Vortices also allow the superfluid component to participate in non-potential
flow and thus in rotation.

The problem of rotation is important in modern physics
\cite{RotatingQuantumVacuum}. It goes back at least to Newton and his
bucket experiment: In a container which rotates at constant angular
velocity $\Omega$ with respect to the inertial frame a viscous liquid
finally relaxes to the equilibrium state -- the state of solid-body
rotation with the container. This state is distinguished eg. by the
parabolic shape of the meniscus, which allows to select the absolute
reference frame in which the surface of the liquid is flat. The superfluid
provides new insights to the problem of rotation: At $T = 0$ (where the
normal component is absent) a vortex-free superfluid remains at rest in the
laboratory frame, its meniscus is flat and thus does not discriminate
between different rotating frames. As superfluids have been found to
simulate in some cases the quantum vacuum -- the modern ether
\cite{Analogue Gravity,Schutzhold2009}, one can ask is rotation an absolute
effect in the quantum vacuum \cite{RotatingQuantumVacuum}. This reminds us
about the arguments of Mach, who claimed that rotation is a relative effect
and can be distinguished only due to coupling to another reference frame
(distant stars).

Quantized vorticity changes the situation: by acquiring an array of
rectilinear vortex lines the superfluid component is able to participate in the
rotational flow and to mimic solid-body rotation. The friction force
between vortices and the normal component provides
the coupling between the container frame and that of the rotating
superfluid, so that in complete equilibrium these frames coincide:
vortices are at rest in the container frame. However, the coupling between
the frames decreases with decreasing temperature, and is
lost in the $T\rightarrow 0$ limit when the normal component vanishes.
Here we discuss an experimental example of the decoupling of the
frames -- the propagating vortex front in
superfluid $^3$He-B.  We find that in the low-$T$ limit, the superfluid
component behind the front develops its own rotating reference frame, whose
angular velocity is smaller than that of the container and is decreasing
with decreasing coupling.

Turbulent vortex-front motion is ubiquitous in the dynamics of both viscous
fluids \cite{ViscousFronts} and superfluids \cite{SuperFluidFronts}. In
superfluid $^4$He turbulent fronts and plugs are known from pipe flow
driven by thermal counter currents of the normal and superfluid components.
In a long cylinder of superfluid $^3$He-B the spin-up response of the
superfluid component from rest to rotation often takes place as an axially
propagating precessing vortex front \cite{Front}. The motion of the front
is launched after a sudden localized turbulent burst. This can be
engineered to happen at constant angular velocity $\Omega$, if the
superfluid is initially in the metastable vortex-free Landau state and the
turbulent burst is triggered externally. Once formed, the front consists of
a turbulent core, with an axial length comparable to the cylinder radius
$R$, and a quasi-laminar tail in the form of a helically twisted bundle
\cite{TwistedCluster} of vortex lines. The front is one of the few
examples of steady-state turbulent motion in superfluids, which are
presently
available for experimental investigations in the
$T\rightarrow0$ limit. We have performed the first direct
measurement of the heat released by such quantum-turbulent motion.

The difference in the free energy of the vortex-free superfluid and that in
equilibrium rotation is $\pi\rho_{\rm s}R^4\Omega^2/4$ per unit length of
the cylinder of radius $R$. Here $\rho_{\rm s}$ is the density of the
superfluid component and a small correction from the discrete nature of
vortices is neglected. In our measurements the spin-up process happens
at $\Omega = {\rm const}$ while the normal component is corotating with the
container. In this case the work performed by the rotation drive can be
found using the balance of angular momentum transfer. It can then be shown
that the heat, expected to be dissipated in the spin-up process, is equal
to the change of the energy of the superfluid.

The simplest model of stationary-state front motion is
that of a thin turbulent front to which all dissipation is concentrated and
where the laminar relaxation to rectilinear vortex lines behind the front
is neglected. If the front moves with the velocity $V$, the
dissipation rate is
\begin{equation}\label{dissi1}
\dot{Q} = \frac{\pi\rho_{\rm s}}{4}R^4\Omega^2 V \,.
\end{equation}
For stationary state motion \Eq{dissi1}
gives the maximum possible rate of heat release. At $0.25 \, T_\mathrm{c}$
and below the front velocity was found in Ref.\,\cite{Front} to be given by
$V \approx \alpha_\mathrm{eff} \, \Omega R$, where $\alpha_\mathrm{eff} \sim
0.1$ is a constant, generated by the turbulence in the front. In this case
the signal intensity is $\dot{Q}\propto \Omega^3$ with only weak temperature
dependence.


\begin{figure}[t]
\centerline{\includegraphics[width=0.98\linewidth]{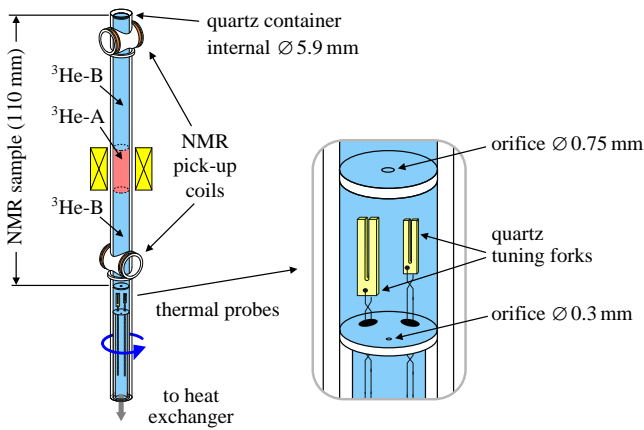}}
\caption{Measuring setup. Two independent cw NMR spectrometers are used to
  monitor vortex front motion in the sample container above the
  $\varnothing$\,0.75\,mm orifice. The middle section with the quartz tuning
  fork oscillators is employed for thermometry and quasiparticle bolometry.
  The bottom section below the $\varnothing$\,0.3\,mm orifice provides the
  thermal contact to the refrigerator. The superconducting solenoid in the
  center of the NMR sample produces the $H_\mathrm{b}$ field for
  stabilizing the $^3$He-A phase. This A-phase barrier layer divides the
  NMR sample in two $^3$He-B sections of equal length.  }
\label{ExpSetUp}
\end{figure}

\textbf{Experimental techniques:}--In Fig.~\ref{ExpSetUp} the $^3$He-B
sample in the top section of the long sample cylinder can be organized to
rotate around its symmetry axis at constant angular velocity $\Omega$ in
the vortex-free state \cite{VorFormAnnih}. Here the superfluid fraction is
at rest in the laboratory frame (${\bf v}_{\rm s} = 0$), while the normal
excitations are in solid-body rotation (${\bf v}_{\rm n}
=\bm{\Omega}\times{\bf r}$). The distribution of vortices and superfluid
counterflow as well as the motion of vortices is surveyed with two NMR
detector coils \cite{ROP}. To trigger the front propagation, we use the
Kelvin-Helmholtz shear flow instability of the AB interface \cite{KH}. The
instability occurs at a well-defined critical velocity $\Omega_{\rm AB}$,
which depends on the magnetic stabilization field $H_\mathrm{b}$ and its gradient at the location of the AB interface. In our
thermal measurements $\Omega_{\rm AB}$ is traversed by sweeping
$H_\mathrm{b}$ at $\Omega = {\rm const}$, for example, by increasing
$H_{\rm b}$ from zero until the A phase is formed. The important consequence from the instability event is the
escape of a number of vortex loops across the AB interface from the A to
the B phase side. The loops interact and produce
a large number of vortices in a turbulent burst close to the AB interface. In the setup of
Fig.~\ref{ExpSetUp} the instability occurs simultaneously at the two AB
interfaces. Thus both upward and downward propagating fronts are set into
motion.

\begin{figure}[t]
\centerline{\includegraphics[width=0.88\linewidth]{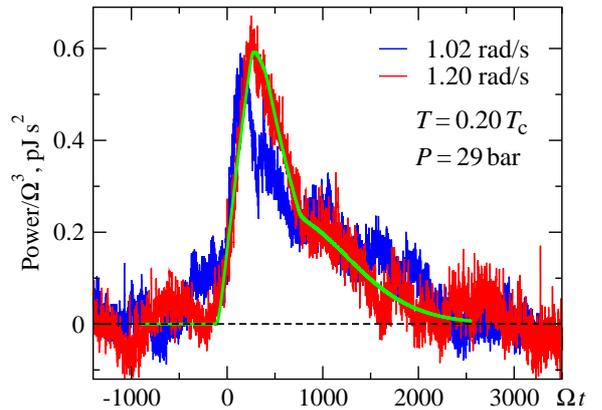}}
\caption{Two $\Omega$-scaled bolometer signals. Time $t=0$ is assigned to
  the moment when the trigger signal (cooling due to the latent heat in B
  to A transition) is observed. This trigger signal and the background heat
  leak has been removed by subtracting the equivalent signal measured in
  the equilibrium vortex state. The thermal time constant of the bolometer
  is about 25\,s, which is much shorter than the time scale of the front
  propagation.  }
\label{FastThermalResponse}
\end{figure}

\textbf{Thermal measurements:}--The heat released in the spin-up process is
recorded as a temperature rise across the lower orifice, which is the
dominant thermal resistance in the ballistic regime. The value of this
resistance is calibrated from the temperature increase as a function of
the known heating power, using one quartz tuning fork oscillator as
thermometer and the other as heater \cite{BBR}. The calibration also gives
a residual heat leak to the $^3$He sample of 15\,pW at $\Omega = 0$, which
increases by $\sim 4\,$pW for a 1\,rad/s increment in $\Omega$.

Examples of the  thermal
measurements of the front propagation are shown in Fig.
\ref{FastThermalResponse}. The total energy release, integrated above
background, matches the expectations: 0.55 and 0.75\,nJ at 1 and
1.2\,rad/s, respectively, versus the expected values of 0.73 and 1.01\,nJ.
The difference of
25\% between the measured and expected heat realease can be ascribed to
uncertainties in the bolometer calibration.

The arrival of the front to the end of the sample, as determined from the
NMR measurements, corresponds to the maximum of the power signal, while
most of the energy is released only after this moment. Thus
by the time the front has reached the end of the sample, the vortex
configuration behind the front still includes less vortices than in the
equilibrium state. The shoulder at $\Omega t\sim 1000$ and the duration of the
signal hint that the late
relaxation behind the front is the laminar spin-up of the superfluid
component \cite {DynamicResponses}.  The fit of the experimental record at
$\Omega = 1.2\,$rad/s to a simple model which
accounts for the turbulent front creating 0.35 of the equilibrium number of
vortices, followed by the laminar relaxation with a time constant of 500 s,
is shown by the green curve in Fig.~\ref{FastThermalResponse} and is in
reasonable agreement with the measurement.

\begin{figure}[t]
\centerline{\includegraphics[width=0.87\linewidth]{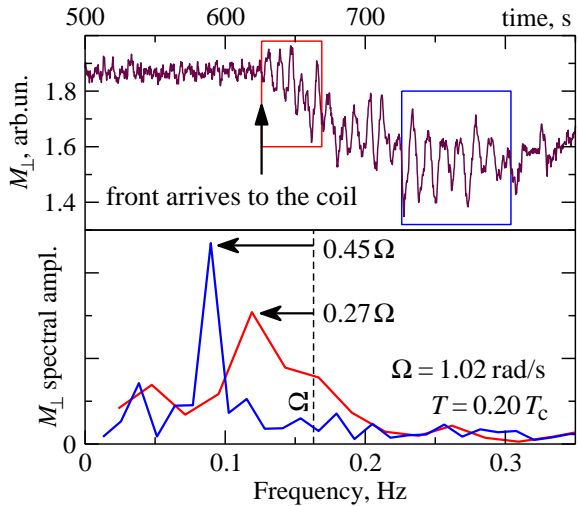}}
\caption{Precessing motion in the vortex front and the vortex bundle behind
  it.  \textit{(Top)} Oscillations in the NMR signal (total transverse
  magnetization $M_\perp$ \cite{BEC-mode}) when the front arrives to the
  pick-up coil result from the precessing motion of vortices.
  \textit{(Bottom)} Fourier transforms of parts of the NMR signal marked in
  the top panel: The front (red) and the cluster behind it (blue). Since
  the NMR coil is fixed to the rotating frame, the precession frequencies
  in the laboratory frame are counted from the angular velocity $\Omega$,
  as shown by the arrows.}
\label{frontosc}
\end{figure}

\textbf{Precessing vortex motion:}--Independent confirmation of the low
density of vortices behind the front comes from observations of
precessing vortex motion around the axis of the cylinder,
Fig.~\ref{frontosc}. The precession is visible as oscillations of the NMR
signal when the vortex arrangement is not perfectly axially symmetric, which
often happens in the front and immediately behind it. For vortex clusters
with uniform density the ratio of their precession frequency $\Omega_{\rm
  s}$ to $\Omega$ is equal
to the ratio of the number of vortices in the cluster to that in the
equilibrium state. For the front itself the precession frequency is about half
of $\Omega_{\rm s}$ \cite{TwistedCluster}. Thus the measurement in Fig.~\ref{frontosc}
shows that at $0.2\,T_{\rm c}$ the cluster behind the front has about 0.4 of
the equilibrium vortex density, in agreement with the thermal signal.

\begin{figure}[t]
\centerline{\includegraphics[width=0.95\linewidth]{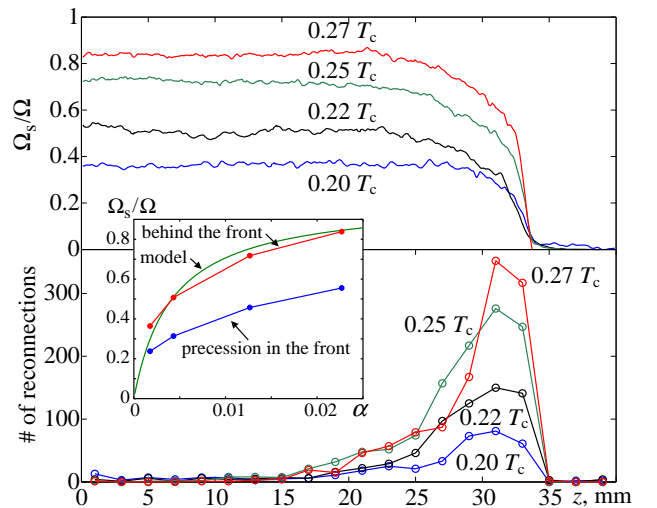}}
\caption{Vortex front in numerical calculations. Snapshots of vortex
  configurations at different temperatures are analyzed when the front is
  located at the same axial position $z=34\,$mm.\textit{(Top)} Azimuthal
  velocity $\langle v_\mathrm{s} \rangle_\phi = \Omega_\mathrm{s} r$ as a
  function of $z$, when averaged over the vortices in the cross section $0
  \leq r \leq 0.8\,R$. The vortex-free state is before the front
  ($z>35\,$mm) and the quasi-equilibrium vortex state with $\Omega_{\rm
    s}(T) < \Omega$ is behind the front. \textit{(Insert)} The values of
  $\Omega_{\rm s}$ behind the front together with a simple model
  Eq.~(\ref{Oms}) and the precession frequency of the vortex ends in the
  front region ($z=30\div35\,$mm) plotted as a function of the mutual
  friction parameter $\alpha(T)$.  \textit{(Bottom)} Turbulence in the
  front is supported by vortex reconnection, which have been counted here
  within bins of width $\Delta z = 2\,$mm over a time interval in which the
  front propagates 0.5\,mm.  Parameters: cylinder length 40\,mm, $R =
  1.5\,$mm, $\Omega = 1.0\,$rad/s, $\alpha (T) = 37.2 \, \exp{(-1.97 \,
    T_\mathrm{c}/T)}$.  }
\label{NumericalResponse}
\end{figure}

\textbf{Numerical calculations:}--To understand the observed reduction in
the number of vortices numerical simulations have been performed using the
vortex filament model and the Biot-Savart law for computing ${\bf v}_{\rm
  s}$ \cite{PLTP-Review}. Above $0.3\,T_\mathrm{c}$, such simulations
demonstrate that the average vortex number per cross section stays constant
as a function of $z$ behind the front and corresponds to solid-body
rotation with the angular velocity $\Omega$ of the container
\cite{PLTP-Review}. Below $0.3\,T_\mathrm{c}$, the vortex density behind
the front also stays constant as a function of $z$, but corresponds to
solid-body rotation with a velocity $\Omega_\mathrm{s} <\Omega$
(Fig.~\ref{NumericalResponse}). In other words, the vortex system develops
its own rotating frame. The velocity of the vortex frame
$\Omega_\mathrm{s}$ decreases prominently with decreasing temperature and
at $0.20\,T_{\rm c}$ agrees with the experimentally determined value of
about $0.4\,\Omega$. $\Omega_{\rm s}$ is found to depend 
weakly on the initial conditions, i.e. on the number and configuration of
the seed vortices which start the front motion. A similar dependence is
observed also experimentally. The calculations confirm that the
precession frequency of the front is approximately $\Omega_\mathrm{s}/2$
(insert in Fig.~\ref{NumericalResponse}). The difference in the rotation
velocities of the front and the vortex bundle behind it is made possible by
vortex reconnections, as seen in the lower panel of
Fig.~\ref{NumericalResponse}. These reconnections lead to the steady-state
turbulence in the front.

\textbf{Decoupling of rotating frames:}--The emergence of a new rotating
frame with angular velocity $\Omega_\mathrm{s}$ can be qualitatively
understood on the basis of the coarse-grained hydrodynamic equation for $ {\bf
  v}_{\rm s}$,
\begin{equation}
\frac{\partial {\bf v}_{\rm s} }{ \partial t}+ \nabla\mu- {\bf v}_{\rm s} \times
(\nabla\times  {\bf v}_{\rm s})  = {\bf F}_{\alpha} + {\bf F}_{\lambda}~.
\label{Hydrodynamics}
\end{equation}
Unlike classical fluids,  two forces appear here on the rhs:  the mutual friction force between the normal and superfluid components,
\begin{equation}
{\bf F}_{\alpha}= - \alpha~\hat{\bf \omega} \times(( {\bf v}_{\rm s}- {\bf v}_{\rm n})  \times
(\nabla\times {\bf v}_{\rm s}) )\,,
\label{MutualFriction}
\end{equation}
and the force due to the line tension \cite{linetens},
\begin{equation}
{\bf F}_{\lambda}= - \lambda (\nabla\times {\bf v}_{\rm s})\times (\nabla\times \hat{\bf \omega})\,.
\label{Tension}
\end{equation}
Here $\hat{\bf \omega} $ is a unit vector along the vorticity $\nabla\times {\bf v}_{\rm s}$,   $\alpha$ is the dissipative mutual friction parameter (while the reactive mutual friction force with the parameter $\alpha'$ is neglected at low $T$).
The line tension parameter $\lambda$ is given by
$\lambda= (\kappa/4\pi)\ln(\ell/a)$,
where $\kappa$ is the circulation quantum, $a$ is the vortex core diameter,
and $\ell$ the inter-vortex distance. Two different Reynolds numbers, $\mathrm{Re}_{\alpha} $ and
$\mathrm{Re}_{\lambda}$,  are
needed to characterize superfluid vortex flow at fixed normal
component. These parameters represent the relative magnitude of the inertial
term in Eq.~(\ref{Hydrodynamics}) with respect to the mutual friction (${\bf F}_{\alpha}$) and line tension (${\bf F}_{\lambda}$) forces:
\begin{equation}
\mathrm{Re}_{\alpha} \approx 1/\alpha~~,~~
\mathrm{Re}_{\lambda}=UR/\lambda = \Omega R^2/\lambda\,.
\label{ReynoldsNumbers}
\end{equation}
The two Reynolds numbers control different regimes of the two fluid
hydrodynamics: for the onset of turbulent flow is required
$\mathrm{Re}_{\lambda}\gg 1$ and $\mathrm{Re}_{\alpha} \gtrsim 1$, while
quasi-classical Kolmogorov turbulence is found in the regime
$\mathrm{Re}_{\alpha} \ < ({\rm Re}_{\lambda})^{1/2}$ and the non-structured
quantum turbulence in the Vinen regime $\mathrm{Re}_{\alpha} >
(\mathrm{Re}_{\lambda})^{1/2}$ \cite{ROP}.

The Reynolds numbers describe the two competing tendencies experienced
by the vortices behind the front: spin-up due to mutual friction from the
normal component, and spin-down by the precessing vortex front, which
rotates slower than the vortex bundle behind the front. The interplay of
these mechanisms leads to the establishment of a quasi-equilibrium
vortex state behind the front. This state is in solid-body rotation
with the velocity $\Omega_{\rm s}$,
which is in general a function of the two Reynolds numbers, changing
from the full coupling ($\Omega_{\rm s} = \Omega$) at high temperatures to
the decoupled state ($\Omega_{\rm s} \rightarrow 0$) in the $T\rightarrow0$
limit. As a simple interpolation between these two regimes we may suggest
\begin{equation}
\Omega_\mathrm{s}/\Omega= (1+  \mathrm{Re}_{\alpha} /
\mathrm{Re}_{\lambda})^{-1},
\label{Oms}
\end{equation}
which is in agreement with the results of the numerical simulations
(Fig.~\ref{NumericalResponse}, insert).

\textbf{Conclusions:}--Both measurements and numerical calculations
demonstrate that the vortex state behind the propagating vortex front
corresponds to quasi-equilibrium solid-body rotation at a velocity
$\Omega_\mathrm{s}$ which is less than the angular velocity of the normal
component: $\Omega_\mathrm{s} < \Omega$. Thus the angular velocity of the
frame in which the vortices are at rest deviates from that of the
container. The decoupling of the two frames becomes more and more prominent
below $0.3 \, T_\mathrm{c}$ with increasing Reynolds number ${\rm
  Re}_\alpha$. This would lead to a total decoupling in the limit $T
\rightarrow 0$, unless a different effect intervenes. Possible candidates
for such zero-temperature coupling include the circular-motion analog of the
Unruh effect \cite{RotatingQuantumVacuum,VolovikRotFriction} and
the interaction of the vortex-core-bound quasiparticle excitations \cite{corebound} with the
container boundaries.


We thank E.B. Sonin for discussions. This work is supported in part by the Academy of Finland (Centers of Excellence Programme  2006-2011 and grant 218211), the EU 7th Framework Programme (FP7/2007-2013,  grant 228464 Microkelvin), and the USA-Israel Binational Science Foundation.




\begin{references}

\bibitem{Volovik2003}
G.E. Volovik,
{\it The Universe in a Helium Droplet},
Clarendon Press,  Oxford (2003).

\bibitem{RotatingQuantumVacuum} P.C.W. Davies \textit{et al.}, Phys. Rev.
  {\bf D~53}, 4382 (1996).

\bibitem{Analogue Gravity}
C. Barcelo \textit{et al.},
Living Rev. Rel. \textbf{8}, 12 (2005).

\bibitem{Schutzhold2009}
R. Sch\"utzhold,
Adv. Sci. Lett. {\bf 2}, 121 (2009).


\bibitem{ViscousFronts} See \textit{eg.} G. Ahlers, D.S. Cannell, Phys. Rev. Lett. \textbf{50}, 1583 (1983); J. Fineberg, V. Steinberg, \textit{ibid.} \textbf{58}, 1332 (1987); B. Hof \textit{et al.}, \textit{ibid.} \textbf{91}, 244502 (2003).

\bibitem{SuperFluidFronts} See \textit{eg.} R.P. Slegtenhorst \textit{et al.}, Physica B \textbf{113}, 367 (1982); and references therein.

\bibitem{Front} V.B. Eltsov \textit{et al.}, Phys. Rev. Lett. \textbf{99}, 265301 (2007).

\bibitem{TwistedCluster} V.B. Eltsov \textit{et al.}, Phys. Rev. Lett. \textbf{96}, 215302 (2006).

\bibitem{VorFormAnnih} V.B. Eltsov \textit{et al.}, J. Low Temp. Phys. \textbf{161}, 474 (2010).

\bibitem{ROP} A.P. Finne \textit{et al.}, Rep. Prog. Phys. \textbf{69}, 3157 (2006).

\bibitem{KH} R. Blaauwgeers \textit{et al.}, Phys. Rev. Lett. \textbf{89}, 155301 (2002).

\bibitem{BBR} S.N. Fisher \textit{et al.}, Phys. Rev. Lett. \textbf{69}, 1073 (1992).

\bibitem{DynamicResponses} V.B. Eltsov \textit{et al.}, Phys. Rev. Lett. \textbf{105}, 125301 (2010).

\bibitem{PLTP-Review} V.B. Eltsov \textit{et al.}, in \textit{Prog.
Low Temp. Phys.} Vol XVI, ed. M. Tsubota (Elsevier B.V., Amsterdam,
2008); preprint arXiv:0803.3225v2.

\bibitem{linetens} R.M. Ostermeyer and W.I. Glaberson, J. Low Temp. Phys.
  \textbf{21}, 191 (1975).

\bibitem{VolovikRotFriction} A. Calogeracos and G.E. Volovik, JETP
  Lett. \textbf{69}, 281 (1999) [Pisma Zh. Exper. Teor. Fiz. \textbf{69},
  257 (1999)].

\bibitem{corebound} M.A. Silaev and G.E. Volovik, J. Low
  Temp. Phys. \textbf{161}, 460 (2010).

\bibitem{BEC-mode} Yu.M. Bunkov \textit{et al.}, preprint arXiv:1002.1674 (2010).




























\end{references}
\end{document}